\documentclass[letterpaper,twocolumn,10pt]{article}
\usepackage{usenix2019_v3}
\usepackage{ulem}
\usepackage{tikz}
\usepackage{amsmath}

\usepackage{filecontents}

\begin{document}
%-------------------------------------------------------------------------------

%don't want date printed
\date{}

\title{\Large{Collective Communication Profiling of Modern-day Machine Learning Workloads}}

%for single author (just remove % characters)
\author{
{\rm Jit Gupta}\\
Juniper Networks
\and
{\rm Andrew Li}\\
Stanford University
\and
{\rm Tarun Banka}\\
Juniper Networks
\and
{\rm Ariel Cohen}\\
Juniper Networks
\and
{\rm T. Sridhar}\\
Juniper Networks
\and
{\rm Raj Yavatkar}\\
Juniper Networks
}
\maketitle

%-------------------------------------------------------------------------------
\begin{abstract}
%-------------------------------------------------------------------------------
Machine Learning jobs, carried out on large number of distributed high-performance systems, involve periodic communication using operations like AllReduce, AllGather, and Broadcast. These operations may create high bandwidth and bursty traffic patterns, leading to network congestion and packet loss, thus impacting the performance of these jobs. Hence it is imperative to analyze these patterns, which can be helpful in provisioning network resources depending on the type of machine learning workloads. In this poster we carry out extensive analysis of the collective communication behavior seen in a wide variety of models (ex. DeepSeek, GPT, Llama, etc.) To achieve this we instrument Nvidia’s Collective Communication Library's logging functionality for richer context about the collectives and workloads. We adjust configuration parameters that influence collective communication behavior, such as parallelism, number of nodes, and model type. This overview presents and discusses some of the results on the collective communication behavior for the open-source DeepSeek V3 inferencing model, which includes operation type and count, transfer sizes per operation, and request size distribution. Our analysis shows that it makes sense to rethink current collective communication frameworks and network topologies so as to accommodate the effect of network anomalies on the mentioned workloads.
\end{abstract} 

\section{Introduction}
Modern day machine learning applications are spread across systems with multiple graphical processing units (GPUs). The workloads can be distributed using parallelism - ex. data parallelism, model and hybrid parallelism, which is a combination of different parallelization techniques\cite{wang2023build}. The training, finetuning or inferencing process requires periodic synchronization between compute nodes (i.e. GPUs), carried out using lossless low latency and high bandwidth network protocols such as RoCEv2\cite{10.1145/2934872.2934908} or Infiniband. 

The communication between compute nodes brings forth the need for collective operations such as AllReduce, AllGather, AllToAll, Broadcast and ReduceScatter. These operations (which determine the GPU-GPU communication) are implemented by frameworks such as NCCL, RCCL, MSCCL, etc.\cite{weingram2023xccl}. It is dictated by the type of application, parallelism and other configuration parameters. However, due to the dynamics of the different operations, traffic patterns and volume of traffic is influenced by aforementioned parameters. This may cause network hot spots leading to congestion and loss, which is be detrimental to the workload's performance. 

\begin{figure}
\centering
  \includegraphics[width=0.7\linewidth]{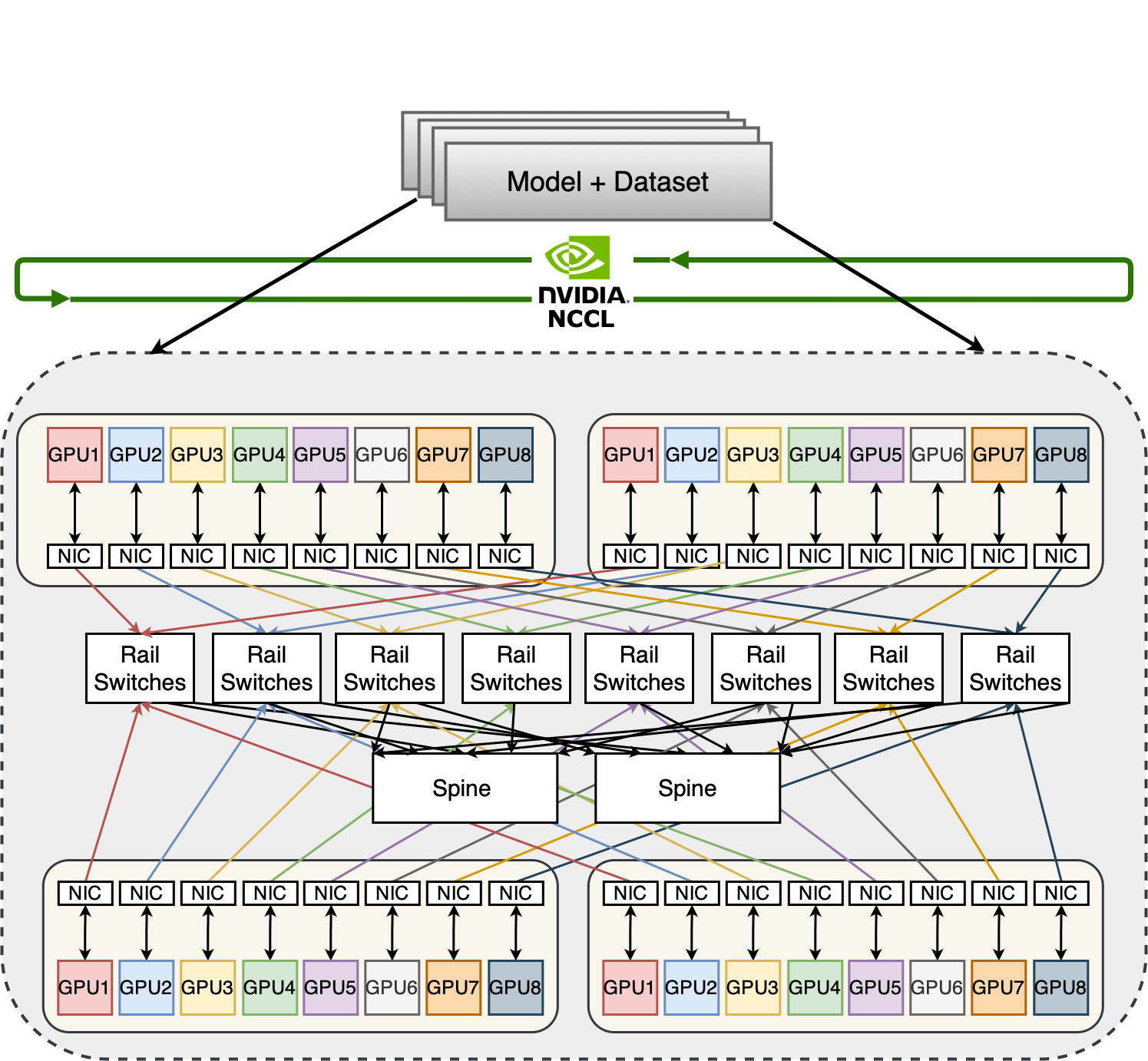}
  \caption{\small Testbed}
  \label{fig:testbed}
  \vspace{-20pt}
\end{figure}

\vspace{-10pt}
\section{Testbed}
We instrument the collective communication library (in this case, NCCL) to log information pertaining to the collective operations. In the current open-source implementation of NCCL, the logging functionality lacks detailed information, which is required to carry out the desired analysis. We log information pertaining to bytes exchanged between GPU pairs (along with the GPUs involved). The GPU-to-GPU field in our logs is used to determine which application is transmitting data from the source GPU to the destination GPU. 
We run DeepSeek V3 (shown in this overview), GPT2, Llama, BERT, Resnet18 and VGG11 using our customized NCCL. The setup contains 4 servers equipped with 8 NVIDIA H100 GPUs (connected using NVlink inside each server) in a rail-optimized topology. The NCCL logs are used to analyze individual workloads.

%-------------------------------------------------------------------------------
\section{Analysis}
\label{sec:analysis}
%-------------------------------------------------------------------------------

\begin{figure}
\centering
  \includegraphics[width=0.4\linewidth]{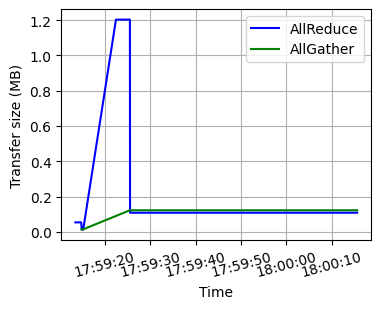}
  \includegraphics[width=0.55\linewidth]{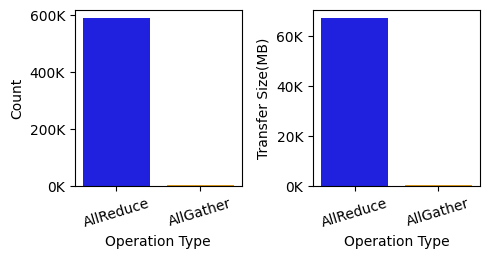}
  \caption{\small(a)Transfer over Time and (b) Operation Comparison}
  \label{fig:transfer}
\end{figure}

We parallelize the DeepSeek v3 inferencing model across 16 GPUs using model parallelism. Following this,, we provide it with eight queries to observe the impact it has on the network. We first observe the type of operations observed, which are AllReduce and AllGather as shown in Fig.\ref{fig:transfer}(a). Training and finetuning workloads also consist of these operations (along with Broadcast), however the transfer sizes seen in those workloads are much larger due to the updation of weights in contrast to inference workloads' activation passing. Additionally, AllReduce operations dominate in all three types of workloads. In the case of DeepSeek we can see that all other operations (i.e. AllGather) are negligible in comparison to AllReduce as shown in Fig.\ref{fig:transfer}(b). The number of AllReduces seen more than $600,000$ for just eight queries, while the number of AllGathers is just $3000$ in comparison.

%\begin{figure}
%\centering
%  \includegraphics[width=0.6\linewidth]{Figures/deepseek_1.png}
%  \caption{Operation Comparison}
%  \label{fig:opscomp}
%\end{figure}

\begin{figure}
\centering
  \includegraphics[width=0.4\linewidth]{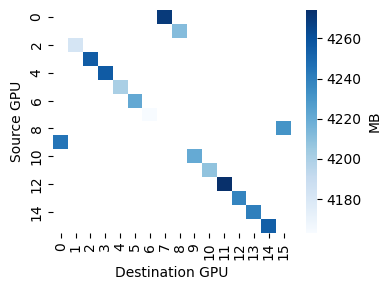}
    \includegraphics[width=0.55\linewidth]{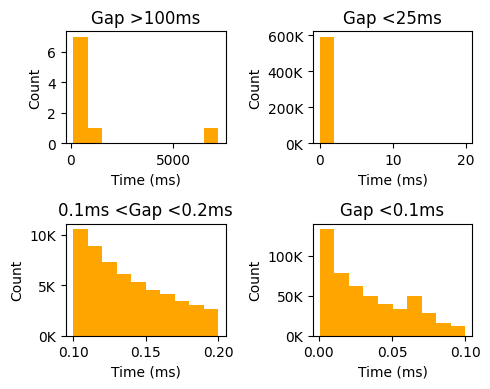}

  \caption{\small (a)Heatmap and (b) distribution of time-gaps}
  \vspace{-20pt}
  \label{fig:heatmap}
\end{figure}

We also observe that though there are a total of 16 GPUs being used, only a handful of them communicate with each other as shown in Fig.\ref{fig:heatmap}. Very little traffic enters the fabric (as each server comprises 8 GPUs) and the spine layer of switches experience negligible traffic. This behavior is also mirrored in training and finetuning workloads. Thus it brings forth the question whether we need a rail optimized topology or if we can modify the topology to accommodate this sparse communication matrix that is shown in Fig.\ref{fig:heatmap}(a).\cite{cassini} 

Analyzing the gaps observed in time per operation is also of importance. This is because datacenter networks are prone to anomalies such as optics failure, port flapping and congestion scenarios. Our analysis for DeepSeek's backend network showed collective communication operations are carried in microsecond granularity (shown in Fig.\ref{fig:heatmap}(a)). Hence in the case of the mentioned anomalous network behavior, the time taken to carry out said operations could extend to tens of seconds. Additionally for training and finetuning workloads, the job itself might fail and might have to be restarted from a previous checkpoint. This is because collective communication frameworks such as NCCL and RCCL do not account for port flapping or optics failure scenarios. Hence it makes sense to incorporate network-aware mechanisms inside them which could gracefully recover from such anomalies.

\section{Conclusion}
This work analyzing the impact of communication between graphical processing units (GPUs) on the network for different types of machine learning workloads. The mentioned communication involves operations called collective communication operations, implemented by frameworks, such as Nvidia’s Collective Communication Library or NCCL. This poster shows detailed analysis of these operations used by different ML workloads i.e., Google’s BERT, OpenAI’s GPT2, Meta’s Llama, DeepSeek, Resnet and VGG. We evaluate effect of different parameters (ex. parallelism, number of nodes, etc.) that determine the traffic pattern due to collective communication. Our results demonstrate variations in network behavior among different types of machine learning training workloads, as well as similarities within ML workloads of the same type (e.g., language and vision models).

%-------------------------------------------------------------------------------

%-------------------------------------------------------------------------------

%-------------------------------------------------------------------------------
\bibliographystyle{plain}
\bibliography{refs}

%%%%%%%%%%%%%%%%%%%%%%%%%%%%%%%%%%%%%%%%%%%%%%%%%%%%%%%%%%%%%%%%%%%%%%%%%%%%%%%%
\end{document}